# Mechanistic Insights Into How Rewiring and Bifurcation Angle Affect DK-Crush Stent Deployment

Andrea Colombo[a], Dario Carbonaro[b], Mingzi Zhang[a,c], Claudio Chiastra[b], Mark Webster[d], Nigel Jepson[e,f], Susann Beier[a]

[a] Sydney Vascular Modelling Group, School of Mechanical and Manufacturing Engineering, University of New South Wales, Sydney, NSW, Australia

[b] Polito<sup>BIO</sup>Med Lab, Department of Mechanical and Aerospace Engineering, Politecnico di Torino, Turin, Italy

[c] Centre for Healthy Futures, Torrens University Australia, Surry Hills, Sydney, Australia

[d] Green Lane Cardiovascular Service, Auckland City Hospital, Auckland, New Zealand

[e] Prince of Wales Clinical School of Medicine, University of New South Wales, Sydney, NSW, Australia

[f] Prince of Wales Hospital, Sydney, NSW, Australia

**Corresponding author:**
Andrea Colombo
School of Mechanical and Manufacturing Engineering, Ainsworth Building (J17), Engineering Rd, University of New South Wales, Kensington, NSW, 2052, Australia
Email: a.colombo@unsw.edu.au
https://orcid.org/0009-0009-6436-6785

**Structured Abstract**

**Background**
Double Kissing Crush (DKC) is a preferred two-stent technique for complex coronary bifurcation lesions. Proximal cell rewiring is routinely recommended to reduce technical failure, and DKC is considered effective across various bifurcation angles. However, it remains unclear whether this standard approach is optimal for all patients. This study investigates the interaction between bifurcation angle and rewiring configuration to identify anatomy-specific strategies.

**Methods**
Computational modeling of the DKC procedure was used to simulate 12 DKC procedures across three left main bifurcation angles (45°, 70°, and 100°) and four rewiring configurations: proximal-proximal (P-P), proximal-distal (P-D), distal-proximal (D-P), and distal-distal (D-D). Evaluation metrics included stent malapposition, side branch ostium clearance, arterial wall stress, low time-averaged endothelial shear stress, and high shear rates.

**Results**
DKC performed in wide bifurcations (100°) resulted in worse outcomes, with malapposition reaching 18%, side branch clearance down to 23%, and up to twice the exposure to adverse high shear rates compared to narrower angles. In contrast, intermediate (70°) and narrow (45°) angles generally resulted in more favorable outcomes, though optimal rewiring varied by angle. Proximal strategies, i.e. P-P and P-D, were most effective at 70°, while D-D performed best at 45°. No single strategy was consistently superior across all bifurcation angles.

**Conclusions**
DKC outcomes depend on bifurcation angle and can be optimized by tailoring rewiring strategies, challenging the current clinical understanding. These findings support anatomy-specific procedural planning and intravascular imaging to guide rewiring. This study provides a mechanistic rationale to improve clinical decision-making and tailor bifurcation interventions.

**Highlights**

- DKC showed suboptimal performance in wide bifurcations (100°)
- Standard P-P rewiring was suboptimal for all bifurcation angles
- Rewiring strategy should be selected based on to bifurcation angle
- Virtual DKC revealed bifurcation angle-technique interaction and supports PCI planning

## 1. INTRODUCTION

Two-stent techniques are widely used to manage complex coronary bifurcation lesions, with the Double Kissing Crush (DKC) technique demonstrating superior outcomes compared to culotte and single-stent techniques [1-4]. Despite its reduced in-stent restenosis and thrombosis rates, DKC remains procedurally complex and operator-dependent, raising concerns about its reproducibility and broader applicability in lower-volume centers [5-7].

In the DKC technique, a double Kissing Balloon Inflation (KBI) step was introduced to improve the success rates of achieving final KBI by adequately reopening the Side Branch (SB) ostium and promoting optimal stent expansion [8]. During this process, both guidewires are typically advanced through proximal stent cells during the rewiring steps [6]. This approach is primarily procedural in nature, aimed at minimizing the risk of wire misplacement outside the stent lumen [9]. Rewiring through distal cells increases the chances of the wire entering between the stent and vessel wall, which may worsen malapposition at the SB ostium and obstruct subsequent re-access with balloons [10, 11]. However, these recommendations are largely procedural and lack biomechanical investigations. Structural Finite Element Analysis (FEA) and Computational Fluid Dynamics (CFD) have been primarily used to optimize stent design, improving the understanding of how stent geometry affects apposition and hemodynamics [12-14]. To date, only one study has modeled the full DKC technique [15], and the role of rewiring configuration has not been investigated. Given the procedural importance of rewiring in DKC, this represents a critical gap in current knowledge.

Clinical studies have reported no significant differences in major adverse cardiac events across bifurcation angles when using the DKC technique [16, 17], suggesting consistent performance in both narrow (< 45°) and wide (> 100°) bifurcation angles. However, these studies are limited to clinical endpoints and do not reveal how bifurcation angle influences the biomechanical behavior of the stented vessel. Computational methods can be used to address this gap, enabling a detailed investigation of how anatomical features interact with procedural steps. *In silico* methods have previously been employed to provide mechanistic insights that complement or refine clinical assumptions [18]. Additionally, other two-stent techniques, such as TAP and culotte, have shown angle dependence: TAP is favored in near-orthogonal angles but may leave the SB ostium uncovered in acute geometries, while culotte tends to perform poorly in wide angles due to excessive strut overlap and incomplete expansion [19]. These observations reinforce the clinical relevance of the bifurcation angle and support the need to examine whether DKC outcomes are similarly influenced.

This study investigates how different rewiring strategies influence the mechanical and hemodynamic outcomes of the DKC technique across bifurcation angles. Our goal was to determine whether a single rewiring strategy can be applied universally or whether its effectiveness depends on bifurcation angle, supporting a more anatomy-specific application of the technique.

## 2. METHODS

### 2.1. DK-Crush Simulations

A population-based coronary bifurcation model was developed with bifurcation branch diameters defined according to Finet's law: 3.25 mm for the Main Vessel (MV), 2.5 mm for the Distal Vessel (DV), and 2.25 mm for the SB [20, 21]. Three bifurcation angles (45°, 70°, and 100°) were selected to represent anatomical variability (Figure 1A). The 70° angle approximates the reported population mean, while 45° and 100° correspond to lower and upper extremes, observed in approximately 10% and 15% of patients, respectively [20]. The arterial wall was modeled as a homogeneous, isotropic, hyperelastic material using a reduced polynomial strain energy function fitted to experimental coronary artery data [22]. The implanted drug-eluting stents featured a geometry resembling the Xience Sierra stent (Abbott Vascular, Abbott Park, IL, USA). Stent diameters and lengths were 2.5 mm and 12 mm for the MV, and 2.25 mm and 8 mm for the SB. The stent material was modeled with an isotropic elasto-plastic formulation to represent the mechanical behavior of a cobalt–chromium alloy [22].

Twelve simulations of the DKC technique were performed by combining the three bifurcation angles with four rewiring strategies: proximal-proximal (P-P), proximal-distal (P-D), distal-proximal (D-P), and distal-distal (D-D). Simulations were conducted using Abaqus/Explicit (Dassault Systèmes, Providence, RI, USA). Each stent was crimped onto a delivery balloon and deployed following established DKC procedural steps [23]. These included SB stenting with minimal protrusion, balloon crush followed by Proximal Optimization Technique (POT) crush, first KBI, MV stenting, first POT, final KBI, and final POT. Balloon diameters were matched 1:1 to the reference vessel, with inflation pressures of 14 atm for KBI inflations, and 18 atm for POT. Rewiring configurations were implemented using the HyperMorph tool in HyperMesh (Altair Engineering, Troy, MI, USA), allowing accurate placement of the balloon through either proximal or distal cells (Figure 1A). The morphing process ensured that the rewired balloon crossed the stent through the designated cell and was aligned correctly within the SB for subsequent expansion. All simulations were conducted under quasi-static conditions, with mass scaling applied to improve computational efficiency. The ratio between kinetic and internal energy was monitored to remain below 5% of internal energy to ensure the quasi-static assumption [24].

Mechanical performance was assessed using three metrics: (1) stent malapposition was quantified as the percentage of struts located more than one strut thickness away from the arterial wall, relative to the total strut surface area; (2) SB ostium clearance was calculated as the ratio between the largest unobstructed area within the SB ostium and the total ostial area, reflecting the degree of metallic obstruction; and (3) arterial wall stress was evaluated by computing both the peak and the average maximum principal stress across all wall elements experiencing stress above 10 kPa, to exclude regions with minimal loading unlikely to elicit a biological response. Although no specific physiological threshold for arterial stress has been defined, elevated mechanical stress promotes vessel injury and may contribute to restenosis through adverse remodeling [25]. In this context, higher stress values are interpreted as indicative of an increased risk of restenotic response.

### 2.2. Hemodynamics Simulations

Hemodynamic simulations were performed subsequently in ANSYS CFX (ANSYS Inc., Canonsburg, PA, USA) using the final stented geometries obtained from the finite element simulations (Figure 1C), following a sequential modeling approach [22]. Blood was modeled as an incompressible, non-Newtonian fluid using the Carreau-Yasuda model [26]. A pulsatile velocity waveform representative of coronary flow was applied at the MV inlet and adjusted in magnitude to reflect the inlet cross-sectional area [27]. Outlet boundary conditions included flow-split at the DV and SB, with flow distributed proportionally according to vessel diameter using allometric scaling laws [28, 29]. A rigid-wall assumption with a no-slip boundary condition at the vessel surface was adopted, as it provides hemodynamic results comparable to those obtained with more complex fluid-structure interaction models in idealized stented coronary arteries [30].

Inlet and outlet extensions were added to ensure a fully developed blood flow across the cardiac cycle [26], and the mesh was refined in the stented region with an element size of 14 µm to achieve an accurate near-wall resolution. Each simulation covered four cardiac cycles, and results were extracted from the final cycle to avoid transient effects [31]. Time step and mesh sensitivity analyses confirmed numerical stability and convergence within a 2% error margin. Two hemodynamic metrics were derived. Adversely high shear rates (> 1000 $s^{-1}$) were evaluated, as they have been associated with platelet activation and increased thrombotic potential [32].

To quantify both the magnitude and extent of adversely high-shear exposure, a shear rate Burden Index (BI) was computed as the product of the blood volume experiencing high shear rates and the average shear rate within that volume. This metric characterizes the overall pro-thrombotic shear environment at peak velocity. Additionally, the percentage of the luminal surface exposed to low Time-Averaged Endothelial Shear Stress (TAESS) (< 0.4 Pa) was calculated to identify regions at risk for inflammation and neointimal proliferation [33, 34]. This metric quantifies the extent of atheroprone regions within the stented bifurcation.

### 3. RESULTS

#### 3.1. Effect of Bifurcation Angle

Bifurcation angle had a marked influence on mechanical and hemodynamic outcomes (Figure 2, Figure 3). Strut malapposition was lowest at 70° (12.0-14.2%), followed by 45° (14.7-15.9%), and worst at 100° for some rewiring configurations (12.5-18.1%), highlighting a trend toward increased malapposition with wide bifurcations. Notably, optimal values for malapposition occurred around the mean bifurcation angle.

SB ostium clearance progressively decreased with increasing angle: at 45°, values ranged from 54-65% across rewiring strategies; at 70°, clearance declined to 36-51%; and at 100°, values dropped further to 23-55% (Figure 4, S1, S2).

Average arterial wall stress showed only modest sensitivity to bifurcation angle, with values ranging from 37-44 kPa at 45°, 26-43 kPa at 70°, and 35-43 kPa at 100°, indicating no systematic trend across angles. Peak arterial stress values were more variable, ranging from 315-1029 kPa at 45°, 241-932 kPa at 70°, and 327-917 kPa at 100° (Table S1).

The high shear rate BI deteriorated substantially with the bifurcation angle. The lowest burden values occurred at 45° (75-140 $mm^3 s^{-1}$), followed by 70° (128-184 $mm^3 s^{-1}$), while 100° exhibited the highest levels of disturbed hemodynamics (124-320 $mm^3 s^{-1}$), driven by increased volume of adversely high shear rate (Table S2). This trend underscores the disruptive influence of wide bifurcations on the shear environment. In contrast, the percentage of luminal surface exposed to low TAESS (< 0.4 Pa) showed limited sensitivity to bifurcation angle, with values ranging from 36.0-41.0% at 45°, 35.1-37.5% at 70°, and 34.9-37.5% at 100°.

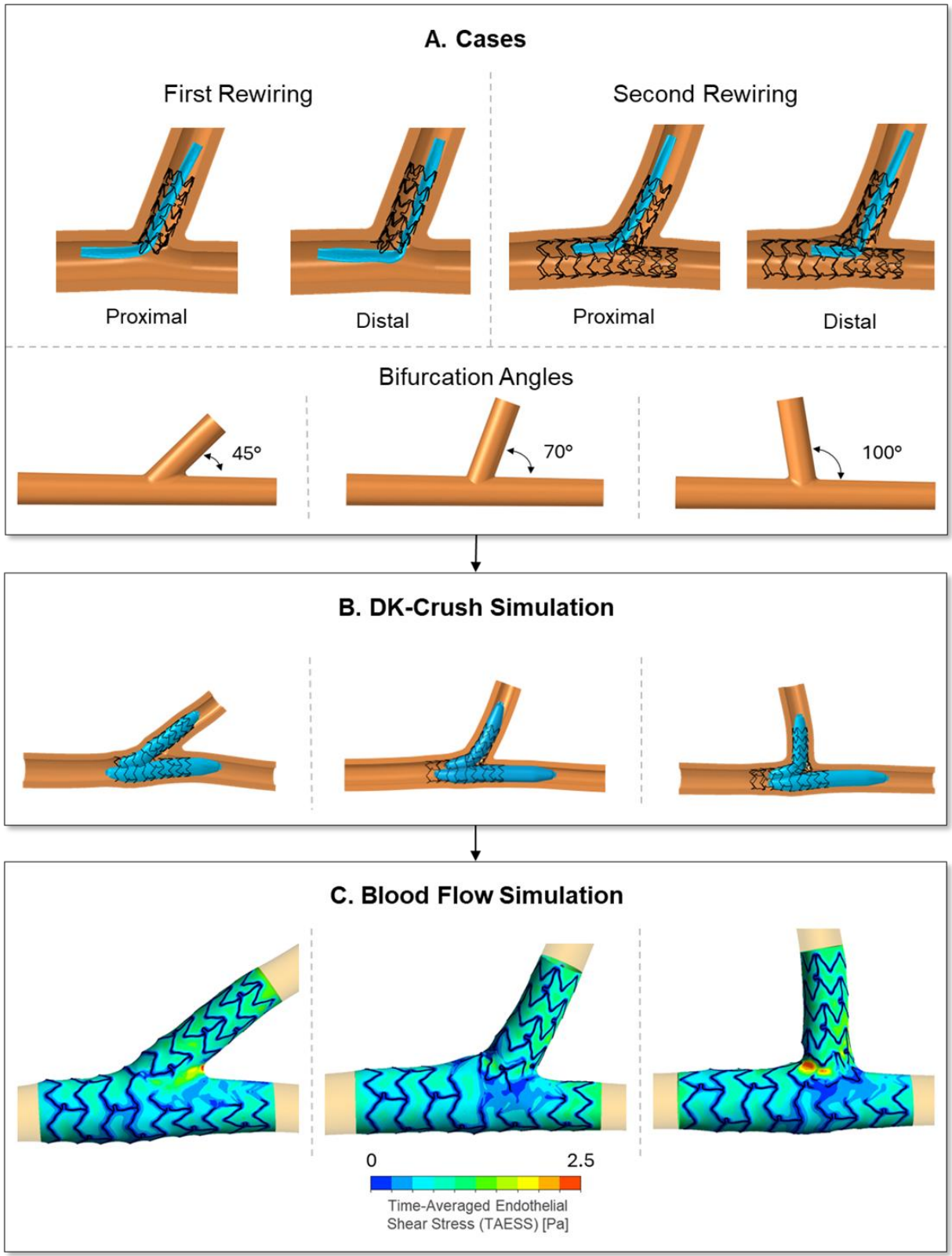

**Figure 1** – Overview of simulation framework used to evaluate the influence of bifurcation angle and rewiring configuration on DKC performance. **A.** The study analyzed 12 combinations of bifurcation angle and rewiring strategy by varying bifurcation angle (45°, 70°, 100°) and guidewire positioning during both the first and second rewiring steps. Each wire was advanced through either a proximal or distal stent cell, resulting in four configurations: proximal-proximal (P-P), proximal-distal (P-D), distal-proximal (D-P), and distal-distal (D-D). **B.** Structural mechanics FEA was conducted to replicate the full sequence of the DKC technique, allowing the quantification of strut malapposition, arterial wall stress, and SB ostium clearance. **C.** CFD simulations were performed on the final deployed configurations

to assess hemodynamic performance, including regions of low TAESS and high shear rate. *CFD: Computational Fluid Dynamics; D: Distal; DKC: Double Kissing Crush; FEA: Finite Element Analysis; P: Proximal; SB: Side Branch; TAESS: Time-Averaged Endothelial Shear Stress.*

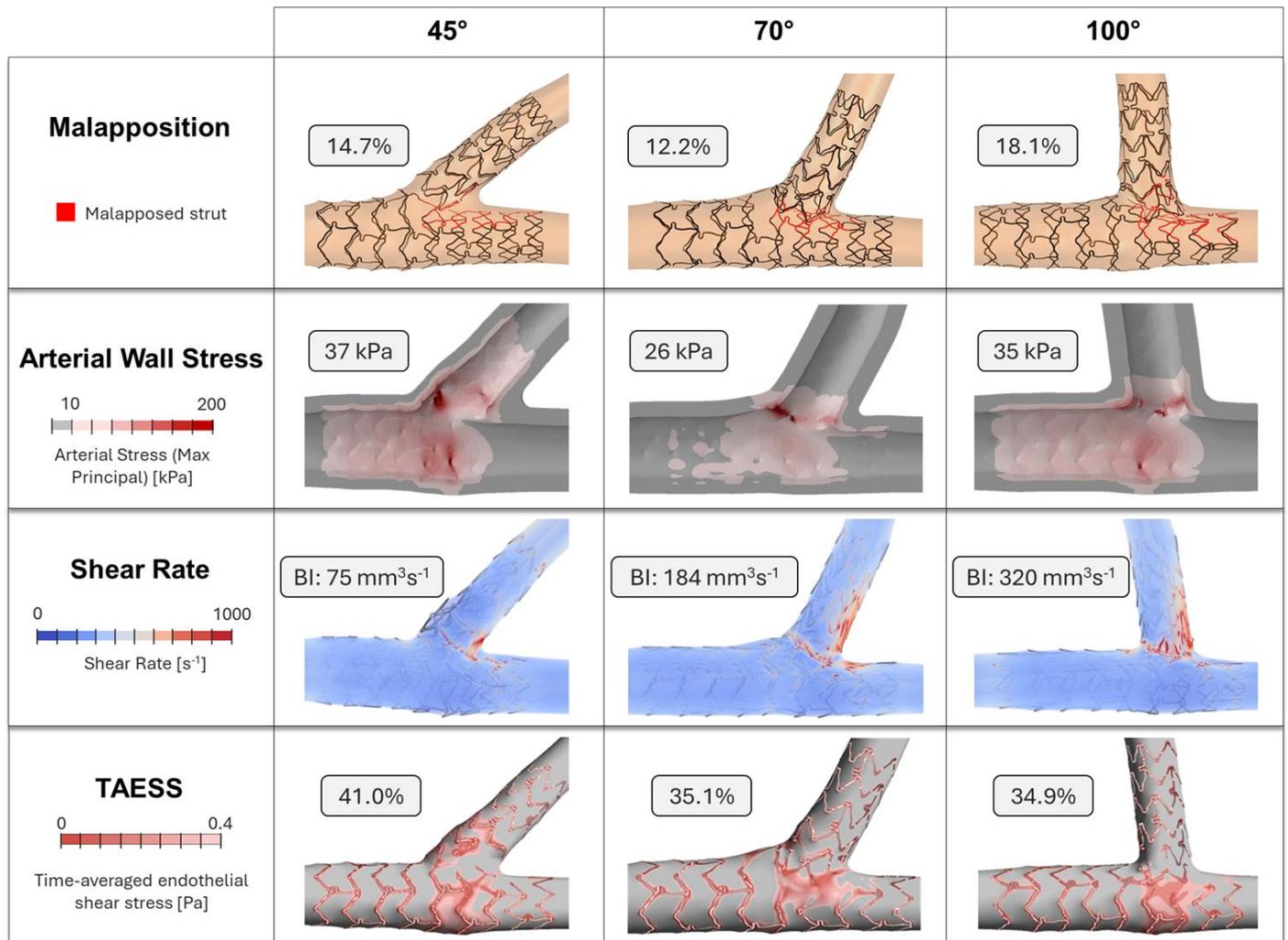

**Figure 2** – Impact of bifurcation angle on key performance metrics of the DKC technique. Representative simulations are shown for the three bifurcation angles (45°, 70°, and 100°). Malapposition (first row) is visualized using P-P rewiring. Average arterial wall stress (second row), shown after D-D rewiring configuration, is shown as a contour plot of maximum principal stress at the end of the DKC stenting. Shear rate (third row) is presented at peak systole, highlighting regions exceeding 1000 s$^{-1}$, with the associated BI values reported. TAESS (bottom row) maps show areas with values below 0.4 Pa, with percentages indicating the relative stented surface area exposed. *BI: Burden Index; Distal; DKC: Double Kissing Crush; P: Proximal; TAESS: Time-Averaged Endothelial Shear Stress.*

### 3.2. Effect of Rewiring Positioning

Rewiring configuration substantially affected SB ostium clearance, arterial wall stress, and shear rate burden (Figure S3). These variations underscore the importance of wire path selection in determining procedural outcome. The following subsections examine each configuration in detail, highlighting its impact across bifurcation angles.

#### 3.2.1. Proximal -Proximal

P-P rewiring was feasible at both 45° and 70° bifurcation angles (Figure 4, Figure S1), where it resulted in the highest SB ostium clearance values (65% at 45°, 51% at 70°), outperforming other configurations by 5-15%. At 100°, however, access was impaired due to the presence of a blocking strut, reducing clearance to 31%. Arterial wall stress was elevated with P-P rewiring, reaching 44 kPa at 45° and 33 kPa at 70°, indicative of increased mechanical load. The shear rate BI exhibited no consistent trend, ranging from 128 mm$^3$ s$^{-1}$ at 70° to 244 mm$^3$ s$^{-1}$ at 100°, highlighting the combined influence of geometry and rewiring on hemodynamics disturbances.

### 3.2.1. Proximal - Distal

At 45°, the second rewiring step was not feasible due to strut crowding following MV stent deployment, resulting in its exclusion from analysis (Figure S1). The configuration was implementable at 70° and 100° but showed lower SB clearance (47% and 23%, respectively), with the 100° case obtaining the worst clearance across all scenarios. Arterial wall stress remained moderate at all angles (32 kPa at 70°, 41 kPa at 100°), and the shear rate BI increased from 148 to 264 mm$^3$ s$^{-1}$ between the two angles.

### 3.2.1. Distal - Proximal

Although rewiring was more technically challenging at 45° and 100°, D-P achieved strong SB ostium clearance at both angles (58% and 55%, respectively) (Figures S1 and S2). In contrast, clearance dropped sharply at 70° (36%), representing the lowest performance for that geometry (Figure 4). Arterial wall stress remained in the upper range (41-43 kPa), particularly at wide angles. This configuration produced the most favorable hemodynamic outcome at 100°, where the shear rate BI was lowest (124 mm$^3$ s$^{-1}$), suggesting improved hemodynamic conditions compared to other strategies.

### 3.2.2. Distal - Distal

D-D rewiring allowed straightforward access at 100°, facilitating procedural completion with lower technical difficulty (Figure S2). Arterial wall stress was consistently low with this configuration, decreasing to 37 kPa at 45°, 26 kPa at 70°, and 35 kPa at 100°, suggesting reduced vessel loading. However, SB ostium clearance was limited across all angles (54-43%), and shear rate BI increased with angle, culminating in the highest value at 100° (320 mm$^3$ s$^{-1}$). These results suggest that D-D may offer procedural simplicity and mechanical protection, but at the cost of compromised ostial access and elevated hemodynamic disturbance in wide-angle anatomies.

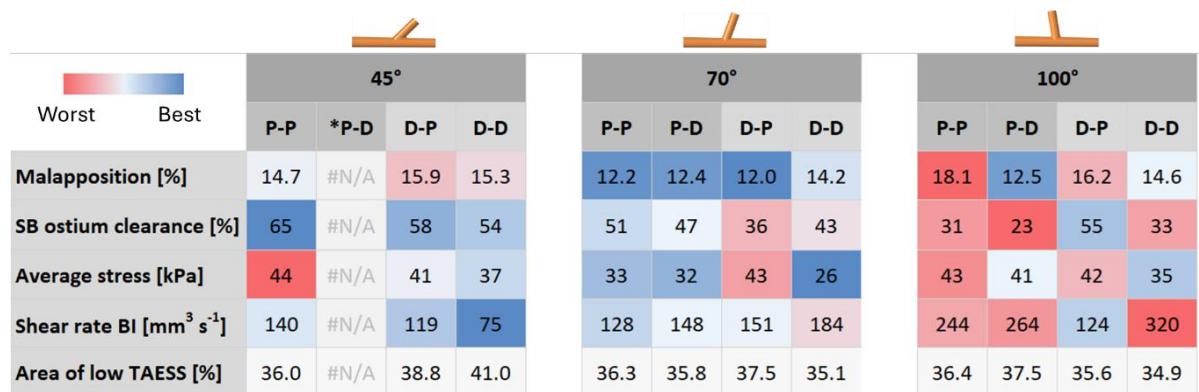

**Figure 3** – Heatmap of performance metrics across bifurcation angles and rewiring strategies. This heatmap summarizes four key mechanical and hemodynamic metrics for each simulated DKC case, combining three bifurcation angles (45°, 70°, and 100°) with four rewiring configurations: proximal-proximal (P-P), proximal-distal (P-D), distal-proximal (D-P), and distal-distal (D-D). Performance metrics include stent malapposition (%), SB ostium clearance (%), average arterial wall stress (kPa), and shear rate BI (mm$^3$s$^{-1}$). Color intensity reflects relative performance, with blue indicating better outcomes and red indicating worse performance for each individual metric.

*The 45° P-D case could not be completed due to strut crowding at the ostium after MV stenting. *BI: Burden Index; D: Distal; DKC: Double Kissing Crush; MV: Main Vessel; P: Proximal; TAESS: Time-Averaged Endothelial Shear Stress.*

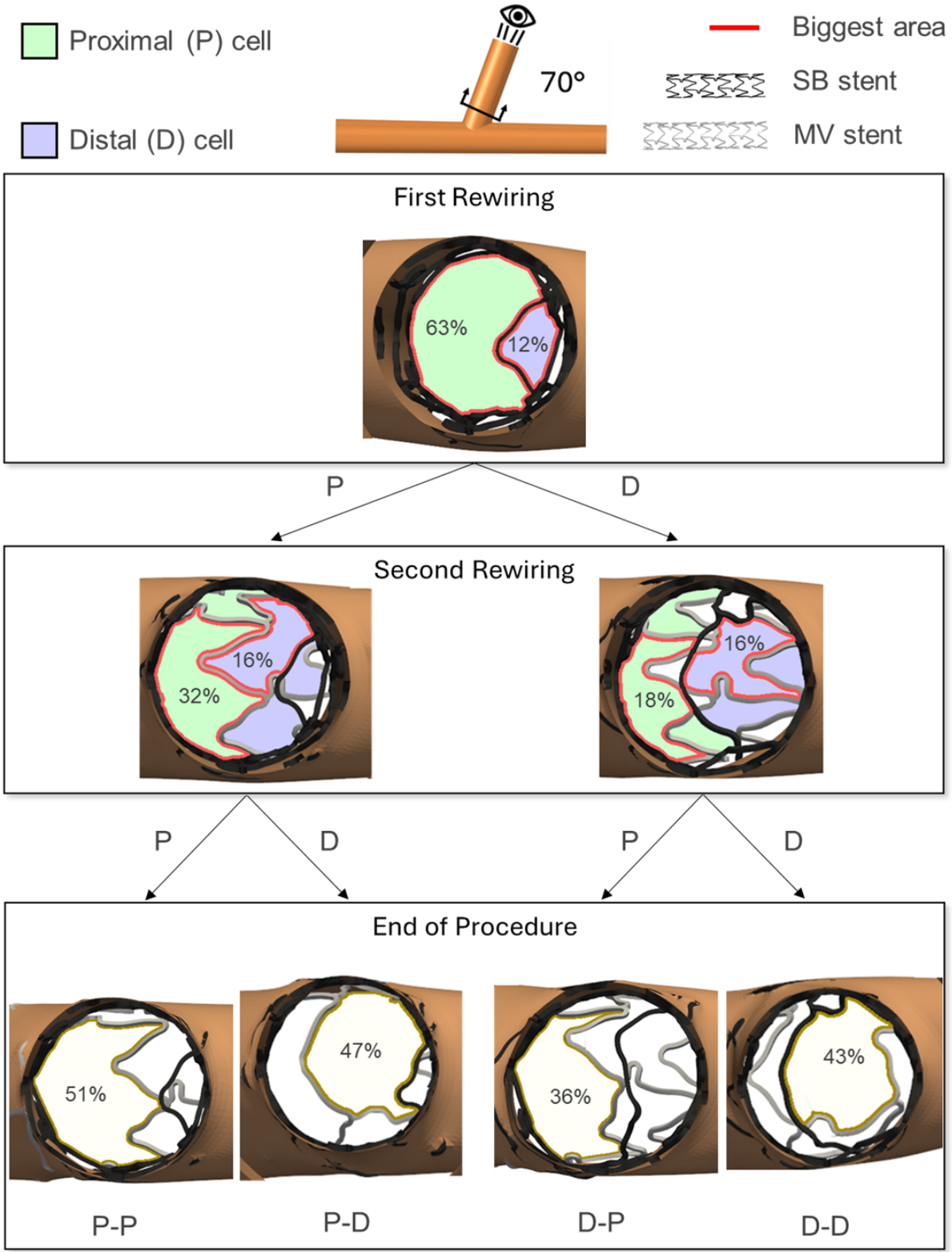

**Figure 4** – Influence of rewiring configuration on SB ostium access and final clearance at 70° bifurcation angle. Cross-sectional views of the SB ostium are shown after each rewiring step and at the end of the DKC procedure. The largest open cell (highlighted in red) and the accessible area through proximal (green) and distal (blue) stent cells are mapped at each stage. MV and SB stents are represented in grey and black, respectively. *D: Distal; DKC: Double Kissing Crush; MV: Main Vessel; P: Proximal; SB: Side Branch.*

## 4. DISCUSSION

This study provides a comprehensive outcome-focused evaluation of DKC technique across different bifurcation angles and rewiring strategies using computational simulations. The findings show that both bifurcation angle and rewiring choice significantly influence mechanical and hemodynamic DKC outcomes that are known to affect clinical treatment success. In particular, a wide bifurcation angle was consistently associated with worse performance across all metrics, and no single rewiring strategy was optimal across all angles. These findings reinforce the need for anatomy-specific DKC planning and challenge the notion of a one-size-fits-all procedural approach.

Although low TAESS is a recognized predictor of restenosis [35], it showed minimal variation across bifurcation angles and rewiring strategies in this study. These findings suggest that TAESS may be more sensitive to other factors, such as stent design or vessel size [15, 36], rather than procedural configuration or bifurcation angle. As a result, TAESS did not offer discriminatory value in this context and was not considered a key determinant of performance in the present analysis.

### 4.1. Effect of Bifurcation Angle

The bifurcation angle emerged as an important determinant of DKC performance. Wide angles (100°) were generally associated with increased malapposition, reduced SB ostium clearance, and elevated shear rate BI, indicating inferior mechanical and hemodynamic outcomes. In contrast, intermediate (70°) and narrow (45°) angles were linked to more favorable performance. At 70°, malapposition was minimized, and ostial clearance remained relatively high across most rewiring strategies. Hemodynamic performance achieved moderate results, suggesting that this angle offers a balance between mechanical performance and flow preservation. Narrow bifurcations (45°) were associated with the lowest shear rate BI and acceptable clearance, although crowding after MV stenting impaired distal access after a first proximal rewiring.

These results challenge the clinical assumption that DKC performs consistently across all bifurcation angles. While clinical trials report no significant differences in angiographic or clinical endpoints between geometries [16, 17], such studies do not capture the underlying biomechanical effects of bifurcation angle on stent deployment. Our results reveal that wide bifurcations were associated with impaired rewiring access, disrupted stent expansion, and disturbed hemodynamics. Although vessel stress varied only modestly across angles, the deterioration in scaffolding and flow metrics suggests that DKC may not be suitable for wide-angle anatomies.

Interestingly, our simulation results align with clinical observations that reported that two-stent techniques, particularly crush, reduce the bifurcation angle by approximately 10° post-stenting, especially in wide bifurcations [37]. This phenomenon was observed in our 100° case, where the angle decreased following stent deployment (Figure 1B).

### 4.2. Effect of Rewiring Positioning

This study also shows that the effectiveness of a given rewiring configuration depends on bifurcation geometry. While the proximal-proximal strategy is employed in clinical practice due to their reduced risk of wire misplacement [9], it does not necessarily achieve the best outcomes in all bifurcation angles.

At 70°, both P-P and P-D configurations facilitated second-wire access, maintained high SB ostium clearance, and kept shear rate BI within moderate ranges. In contrast, P-P performed poorly in wide bifurcations due to obstructed access paths and compromised ostial scaffolding. This resulted in low SB clearance and elevated high shear rate burden. Among all configurations, D-P rewiring yielded the most balanced outcome, suggesting it might partially offset the adverse effect of wide angles.

Conversely, in narrow bifurcations, D-D rewiring offered the most balanced mechanical and hemodynamic performance, with reduced arterial stress and adverse shear rate.

Across all cases, proximal-first rewiring (P-P, P-D) consistently increased arterial stress, suggesting that re-entry through proximal cells may impose greater mechanical load. Conversely, D-D configurations yielded the lowest stress values, potentially offering mechanical advantages in fragile vessels. Importantly, peak arterial wall stresses were consistently concentrated on the lateral wall opposite the carina. This region is often subjected to low shear forces and is prone to plaque development and structural weakening [38]. While malapposition and low TAESS area were relatively unaffected by rewiring, metrics such as SB clearance, shear strain, and arterial stress showed notable sensitivity to wire path.

Previous simulation studies on provisional stenting reported that distal rewiring improved SB scaffolding and ostial coverage, supporting its use in one-stent strategies [39, 40]. In contrast, our DKC simulations revealed that distal rewiring did not consistently result in superior SB ostium clearance. This suggests that optimal rewiring positioning in DKC depends more on the specific patient and procedural complexity, and that findings from simpler one-stent techniques may not directly apply to two-stent approaches.

Final KBI was successfully achieved in 11 out of 12 simulated cases, regardless of angle or rewiring configuration. This confirms that DKC reliably allows for final KBI across a range of anatomies [8]. However, SB access and scaffolding quality varied substantially, reinforcing that successful procedural execution alone does not ensure optimal mechanical or flow-related outcomes.

### 4.3. Clinical Relevance

In clinical settings, operators routinely treat bifurcations that vary considerably across the patient population. Approximately 10% of lesions present with narrow angles (< 45°), and 15% with wide angles (> 100°), with males typically exhibiting wider bifurcations than females (average 85° vs. 74°) [20]. These anatomical differences have direct implications for procedural planning and rewiring strategy selection in DKC interventions.

In wide bifurcations, DKC was associated with impaired rewiring access, poor SB ostium scaffolding, and increased biomechanical burden. These results suggest that DKC may be suboptimal in these patients and alternative stenting techniques should be considered where feasible. However, if DKC is selected based on the operator's preference, D-P rewiring may mitigate some adverse outcomes by improving SB access and reducing hemodynamics disturbances compared to other configurations.

For bifurcation angles near the population average (70°), proximal-first strategies such as P-P and P-D were associated with favorable outcomes, including optimal rewiring access, good ostial expansion, moderate shear rates and balanced wall stress, reinforcing the current procedural preference for proximal cell access based also on outcomes.

In narrow bifurcations (< 45°), where post-stenting strut crowding and angulation might hinder rewiring, the D-D strategy offered the most balanced performance. D-D minimized arterial wall stress while maintaining ostial patency, making it a potentially advantageous choice when treating acute-angle bifurcations or lesions with limited proximal access.

Overall, the findings on rewiring positioning reinforce the importance of intravascular imaging, particularly Optical Coherence Tomography (OCT), in guiding bifurcation stenting. OCT allows precise identification of rewiring entry points, allowing operators to optimize cell selection and minimize incomplete stent apposition. Our simulations demonstrate that rewiring strategy directly impacts mechanical and hemodynamic outcomes, and that distal or alternative cell entry may offer advantages over standard proximal approaches in certain patients. Together, these results support the routine use of imaging to tailor guidewire positioning and enhance procedural precision, especially in wide or complex bifurcations [38, 41].

### 4.4. Limitations

This study employed population-based vessel geometries and uniform material properties to isolate the influence of bifurcation angle and rewiring strategy on DKC outcomes. While this approach excludes the variability introduced by patient-specific anatomy and plaque heterogeneity, it improves reproducibility and enables clearer mechanistic comparisons across procedural configurations [42, 43]. The absence of calcification or fibrotic plaque modeling may underrepresent the mechanical challenges encountered in clinical practice. In particular, the presence of disease could worsen strut malapposition, wall stress, and shear rates [44]. Procedural performance might also be affected, as wire recrossing could be mechanically constrained. However, the idealized conditions allow for a robust baseline evaluation of stent-vessel interaction.

This study used only one stent design, specifically a geometry resembling the Xience Sierra stent, which is among the most implanted stents in coronary interventions. Although this choice supports clinical relevance, different stent platforms may yield different outcomes due to variations in ring design and material properties, which can affect strut distribution at the ostium and ease of rewiring. These factors may lead to procedural differences not captured by the current model and should be explored in future comparative studies.

Rewiring paths were predefined as proximal or distal based on geometric location, whereas in clinical procedures, wire positioning may be influenced by imaging modalities or operator judgment. Furthermore, the study examined a limited set of configurations, three bifurcation angles and four rewiring strategies, chosen to represent clinically relevant extremes and common procedural options. While these do not include the full anatomical and procedural spectrum, they provide a representative framework for evaluating the performance range of DKC. The results suggest that the relationship between bifurcation angle and stenting outcomes is non-linear, with intermediate angles associated to better performance and wider or narrower angles introducing distinct challenges. Future studies incorporating patient-specific geometries, lesion morphology, and material heterogeneity could offer more personalized insights but would significantly increase model complexity and computational cost. Nonetheless, the current findings offer a strong foundation for understanding key anatomical-procedural interactions in complex bifurcation stenting.

### 5. CONCLUSION

This study demonstrates that the bifurcation angle directly impacts the performance of the DKC technique and should be a primary consideration during procedural planning. Wide bifurcations

(>100°) were associated with reduced SB ostium clearance, increased malapposition, higher shear strain burden, and greater difficulty in rewiring access, leading to inferior overall outcomes. These findings challenge the assumption that DKC yields consistent results across bifurcation geometries. Rewiring configuration also critically influenced procedural performance, but its effectiveness was dependent on the bifurcation angle. While proximal-first strategies (P-P, P-D) performed well in intermediate angles, D-D and D-P were superior in narrow and wide-angle anatomies, respectively. The commonly used P-P approach did not consistently yield the best outcomes, particularly in wide bifurcations. These findings support adapting a rewiring strategy to bifurcation geometry rather than adopting a fixed procedural sequence. Computational modeling may assist in guiding these choices by identifying anatomically tailored configurations that optimize stent deployment and flow restoration.

## 6. Declaration of generative AI and AI-assisted technologies in the writing process

During the preparation of this work the author(s) used ChatGPT and Grammarly to improve the readability and language of the manuscript. After using this tool/service, the author(s) reviewed and edited the content as needed and take(s) full responsibility for the content of the published article.

**Supplementary Material**

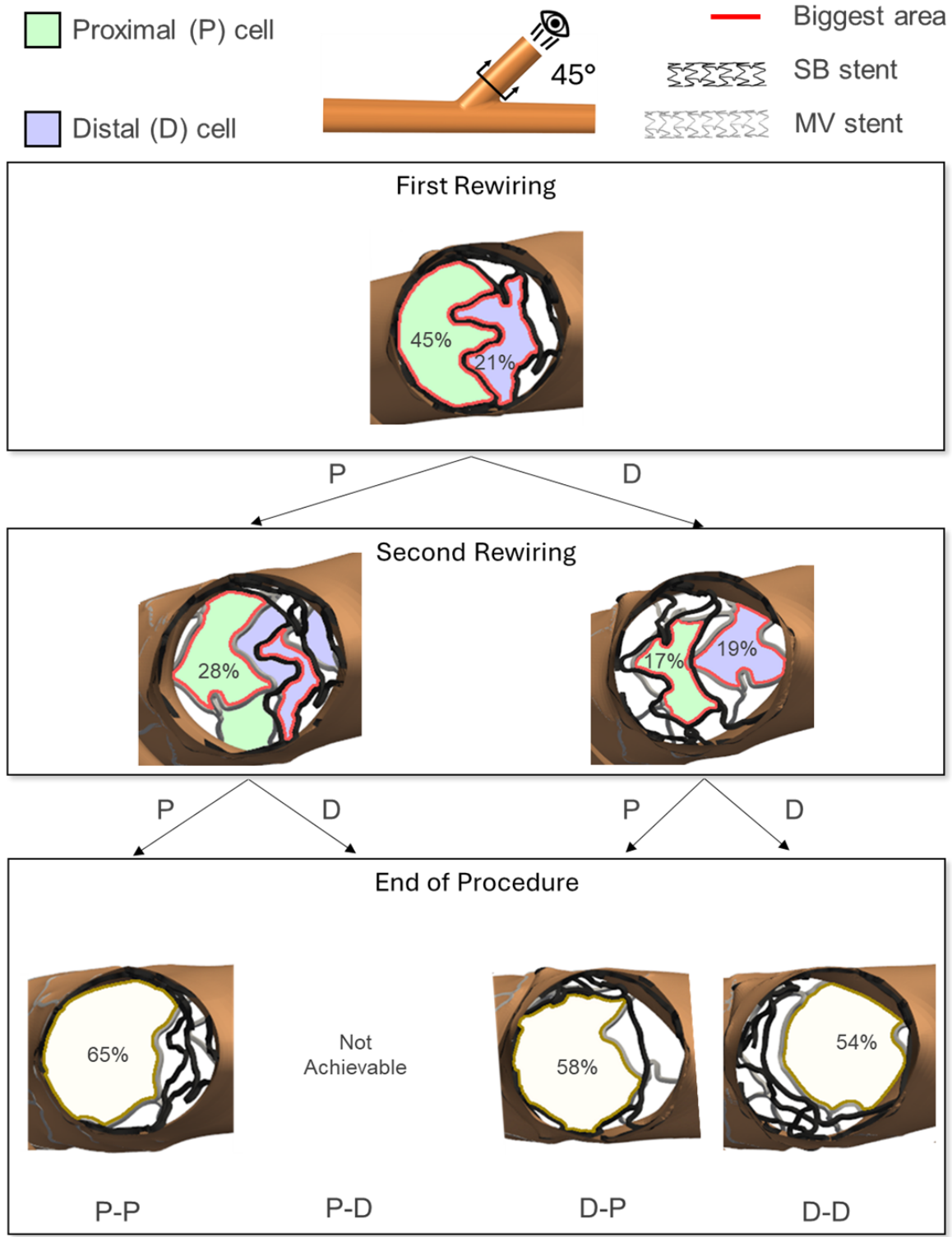

**Figure S5** – Influence of rewiring configuration on stent cell selection and SB ostium clearance in a 45° bifurcation. Cross-sectional views illustrate the selection of proximal (green) and distal (purple) stent cells during first and second rewiring steps of DKC, with the largest accessible cell outlined in red. The MV and SB stents are shown in grey and black, respectively. P-D was not achievable due to limited access after MV stent deployment. *D: Distal; DKC: Double Kissing Crush; MV: Main Vessel; P: Proximal; SB: Side Branch.*

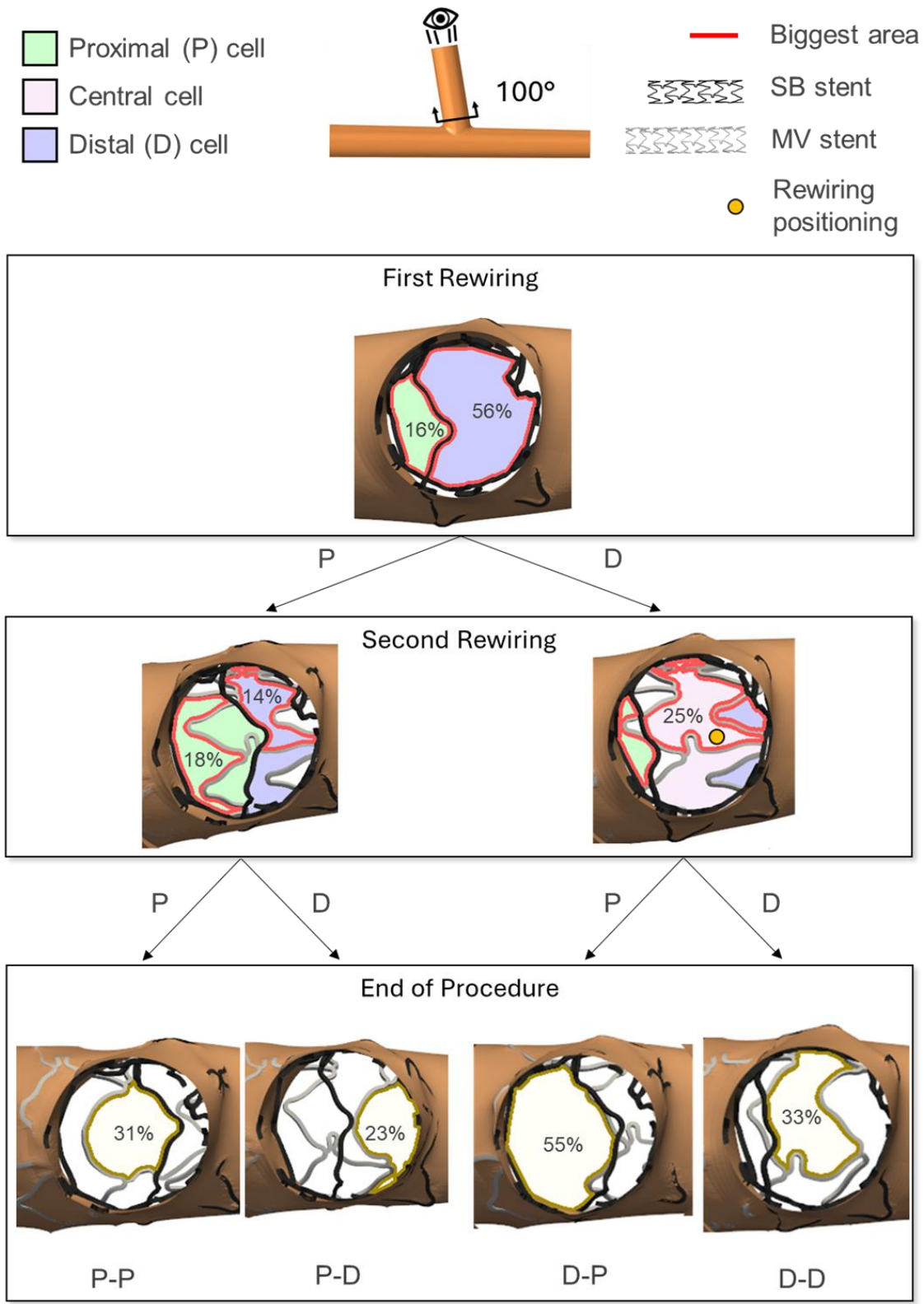

**Figure S6** – Influence of rewiring configuration on stent cell selection and SB ostium clearance in a 100° bifurcation. Cross-sectional views illustrate the selection of accessible stent cells at the SB ostium during the first and second rewiring steps of the DKC technique. Proximal (green), central (pink), and distal (blue) cells are mapped, with the largest accessible cell outlined in red. The MV and SB stents are shown in grey and black, respectively. The yellow dot indicates the actual position of rewiring through a central MV stent cell. Final ostial clearance for each rewiring configuration was defined as the percentage of SB ostium area unobstructed by stent struts at the end of the procedure. MV stent. *D: Distal; DKC: Double Kissing Crush; MV: Main Vessel; P: Proximal; SB: Side Branch*.

| | P-P | | | P-D | | | D-P | | | D-D | | |
|---|---|---|---|---|---|---|---|---|---|---|---|---|
| | 45° | 70° | 100° | *45° | 70° | 100° | 45° | 70° | 100° | 45° | 70° | 100° |
| Malapposition [%] | 14.7 | 12.2 | 18.1 | #N/A | 12.4 | 12.5 | 15.9 | 12.0 | 16.2 | 15.3 | 14.2 | 14.6 |
| SB ostium clearance [%] | 65 | 51 | 31 | #N/A | 47 | 23 | 58 | 36 | 55 | 54 | 43 | 33 |
| Average stress [kPa] | 44 | 33 | 43 | #N/A | 32 | 41 | 41 | 43 | 42 | 37 | 26 | 35 |
| Shear rate BI [mm³ s⁻¹] | 140 | 128 | 244 | #N/A | 148 | 264 | 119 | 151 | 124 | 75 | 184 | 320 |
| Area of low TAESS [%] | 36.0 | 36.3 | 36.4 | #N/A | 35.8 | 37.5 | 38.8 | 37.5 | 35.6 | 41.0 | 35.1 | 34.9 |

**Figure S7** – Influence of rewiring strategy on DKC performance across bifurcation angles. Heatmaps report four key outcome metrics, malapposition, SB ostium clearance, average arterial wall stress, and shear strain burden index, for each combination of rewiring strategy (P-P, P-D, D-P, D-D) and bifurcation angle (45°, 70°, 100°). Values are color-coded from blue (favorable) to red (unfavorable). *The 45° P-D case could not be completed due to procedural limitations. *DKC: Double Kissing Crush; SB: Side Branch*.

**Table S1** – Mechanical results for the 12 simulated cases, by varying bifurcation angle and rewiring configuration after double kissing crush. *D: Distal; P: Proximal; SB: Side Branch. TAESS: Time-Averaged Wall Shear Stress.*

|  | **45°** | | | | **70°** | | | | **100°** | | | |
| --- | --- | --- | --- | --- | --- | --- | --- | --- | --- | --- | --- | --- |
| **Stent Deployment** | P-P | P-D | D-P | D-D | P-P | P-D | D-P | D-D | P-P | P-D | D-P | D-D |
| Malapposition (%) | 14.7 | - | 15.9 | 15.3 | 12.2 | 12.4 | 12.0 | 14.2 | 18.1 | 12.5 | 16.2 | 14.6 |
| SB ostium clearance (%) | 64.5 | - | 58.1 | 53.9 | 51.7 | 46.6 | 36.1 | 43.2 | 31.3 | 22.8 | 55.1 | 32.9 |
| Max artery stress @ end [kPa] | 1029 | - | 605 | 315 | 533 | 642 | 932 | 241 | 621 | 654 | 917 | 327 |
| Average artery stress @ end [kPa] | 44 | - | 41 | 37 | 33 | 32 | 43 | 26 | 43 | 41 | 42 | 35 |

Table S2 – Hemodynamics results for the 12 simulated cases, by varying bifurcation angle and rewiring configuration after double kissing crush. D: Distal; P: Proximal; SB: Side Branch; TAESS: Time-Averaged Wall Shear Stress

| | *45°* | | | | *70°* | | | | *100°* | | | |
|---|---|---|---|---|---|---|---|---|---|---|---|---|
| *Blood Flow* | P-P | P-D | D-P | D-D | P-P | P-D | D-P | D-D | P-P | P-D | D-P | D-D |
| TAESS [Pa] | 0.48 | - | 0.48 | 0.49 | 0.47 | 0.49 | 0.48 | 0.51 | 0.50 | 0.51 | 0.50 | 0.53 |
| % area with TAESS < 0.4 Pa | 36.0 | - | 38.8 | 41.0 | 36.3 | 35.8 | 37.5 | 35.1 | 36.4 | 37.5 | 35.6 | 34.9 |
| Volume with shear rate > 1000 $s^{-1}$ [$mm^3$] | 0.09 | - | 0.08 | 0.05 | 0.08 | 0.10 | 0.10 | 0.13 | 0.16 | 0.18 | 0.09 | 0.22 |
| Average shear rate > 1000 $s^{-1}$ [$s^{-1}$] | 1498 | - | 1490 | 1382 | 1527 | 1458 | 1531 | 1434 | 1512 | 1460 | 1427 | 1482 |
| Shear rate burden index [$mm^3 s^{-1}$] | 140 | - | 119 | 75 | 128 | 148 | 151 | 184 | 244 | 264 | 124 | 320 |